# An Early Development of Flood Inundation Mapping Method Using Free Android Application to Support Emergency Response Activities: Case Study Baleendah, Bandung, West Java

*Sandy Hardian Susanto* Herho[1*], *Edi* Riawan[1], *Cahya* Nugraha[1], *Rusmawan* Suwarman[1], *I Dewa Gede Agung* Junnaedhi[1], and *Dasapta Erwin* Irawan[2]

[1]Atmospheric Science Research Group, Faculty of Earth Sciences and Technology, Bandung Institute of Technology (ITB), Indonesia

[2]Applied Geology Research Group, Faculty of Earth Sciences and Technology, Bandung Institute of Technology (ITB), Indonesia

**Abstract.** This study aims to simplify the current flood mapping methodology, so it can be done in quicker time, easy to learn, but still produce accurate flood information to support the emergency responses. This study is divided into two activities: Survey development and flood mapping methods; Field survey and mapping to test the developed method. The method development was then conducted by examining the simple methods in the flood mapping survey and then testing it in the field. The results of these trials were then evaluated to determine the most effective and efficient methods. The utilization of free android mapping application to conduct flood survey shows satisfactory results. Based on two trials, it was known that it takes only 4 hours to conduct a 15 km survey of Citarum River segment. The mapping shows that the flood areas in Baleendah and its surrounding reached 763 Ha and 794 Ha respectively on 25 December 2014 and 13 March 2016. In addition, the method developed is also relatively easy to use, so it is expected to trigger the local communities to play an active role in disaster prevention efforts, especially in emergency response by providing accurate information about the flood inundation areas.

## 1 Introduction

Flood is one of the most destructive natural disasters for humans and the others life around it (Bellos, 2012). Flood generally occurs in low areas with flat topography and is adjacent to the medium of water flow, such as trenches and/or rivers. Flood occurs when the medium of water flow is not able to accommodate the volume of flow. The consequence is water overflows to the surrounding area into a puddle of flood. Causes of flood include natural events and human activities, and it may be a combination between these two causes. The natural events that caused the occurrence of flood disasters include high rainfall, inadequate river flow capacity, and river flows retained by river flow (Hapit, 2014). Human activities that resulted in flood include the development of areas along the banks of river basins, changes in land use in watersheds that caused larger run-offs, river banks utilized as settlements, and lack of public awareness to dispose of waste in its own place (Santoso, 2012).

Bandung basin is part of the Upper Citarum Basin. Hydrologically the Upper Citarum watershed experienced considerable degradation and was categorized as a critical watershed in Indonesia (Dasanto et al, 2014). In the rainy season, the flow of water at Citarum watershed is very high, causing annual floods that hit several sub-districts along the river, especially in low areas such as Baleendah Subdistrict, Bandung Regency. Major floods caused by overflows of the Citarum river that occurred in Baleendah and surrounding areas were recorded in 1986, 1998, 2005 and 2010 (BBWSC, 2011). These incidents cause considerable losses that could disrupt the welfare of local communities. To overcome the frequent flood problems at Baleendah, good flood risk management involving local communities is required. One such flood risk management effort is the creation of a flood inundation map conducted by local residents for emergency response of the flood-prone region.

Flood inundation map is a topographic map that showing the extent of the area which affected by flood water in a flood event (Merwade et al, 2008). There are three main methods that commonly used for flood inundation mapping, using numerical model; physical model; and field survey method (historical record method) (Bellos, 2012). This study was conducted by piloting the field survey method using a free Android-based mapping application that was done by the local community as an independent emergency response effort.

[*] Corresponding author: sandyherho@meteo.itb.ac.id

## 2 Data and Methods

The primary data that has been used in making of the early flood inundation map in Baleendah are the coordinate data of flood inundation reach, that was collected by the survey team from the local residents. Collection of flood boundary coordinates is done by using an Android-based free mapping application, Orux Map (http://www.oruxmaps.com/cs/en). This survey was conducted twice, i.e. flood incident on 25 December 2014 and 13 March 2016. The survey zone is divided into two parts, namely the northern zone and the southern zone bounded by the Citarum river. The division of this survey zone is based on informal information from local residents and BPBD about the farthest flood reach point. Each zone is divided into five entry points based on the availability of roads that can be taken by the survey team motorcycles, for the capture of the flood reach point coordinates (**Fig 1**). The survey team consist of 13 surveyor personnel who are divided into six teams, with the placement of five teams at the northern zone and one team at the southern zone.

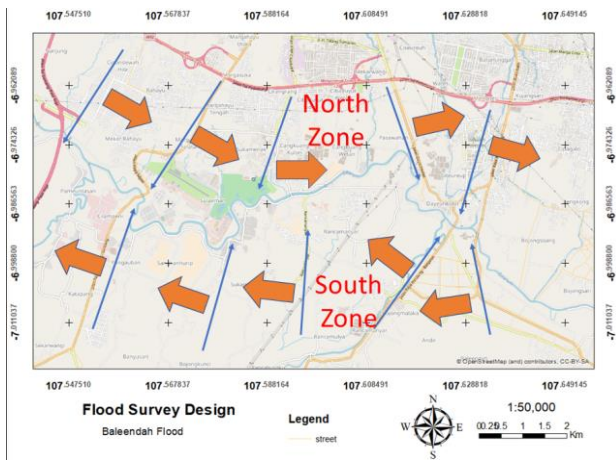

**Fig. 1.** Design of rapid flood mapping survey. The blue arrow indicates the movement of the motorcycle used by the survey team in the alley towards the river for plotting the reach of flood inundated area, and the orange arrow indicates the movement of the team towards the next alley.

This survey produce two pieces of information. First, the coordinates of the reach point of the flooded immersion accompanying the photo of the immersion limit, wherein this information indicates the boundaries of the flood to be connected to the line as the boundary of the inundation area. Second, surveyors path point to areas that are less likely to be flooded. Flood inundation reach coordinate points that has been obtained from this survey, processed digitally with the help of flood expert team from WCPL ITB. The flood inundation maps were estimated by connecting the coordinate points of flood reach survey results and taking into account the topographic data (DEM, Terrasar).

## 3 Results and Discussion

Two inundation flood mapping trials were conducted on 25 December 2014 and 13 March 2016. In the first mapping experiment (**Fig.2a**) the surveyor took 100 coordinate points. The mapping result showed that the flooded area reached 763,48 Ha. The Dayeuhkolot Village is the most affected village with the most severe flooding area covering 93.15% of the total area of the village. The village with the widest flooded area in Baleendah Subdistrict on 25 December 2014 is Andir village with an inundated area of 211,16 Ha. In the second mapping test (**Fig.2b**), the points that taken by the surveyor team was 69 points, with the mapping result indicating the area of the flooded area reached 794.1 Ha. The flood inundation pattern of both events shows similarities.

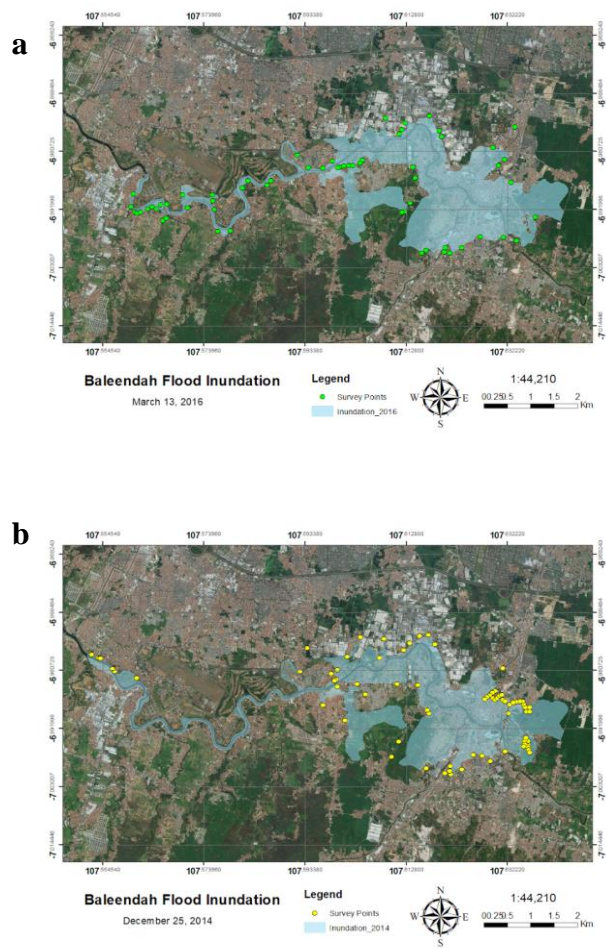

**Fig. 2.** Baleendah Flood Inundation Map on (a) 25 December 2014 and (b) 13 March 2016.

The reliability of this rapid flood survey method is evident from its speed and accuracy. Based on two flood mapping trials, the average survey rate for each team is up to 6.2 km/h using the free Android-based mapping application, **Orux Map**. The more specific result can be seen in Table 1. To support emergency response efforts, it is necessary to speed up the construction of flood inundation maps. With this method, in addition to the data processing team at the post, the team in the field can also perform real-time data processing through plotting with Google Earth application (**Fig.3**), thereby speeding up the evacuation of local residents while the surveyors team conducting field surveys.

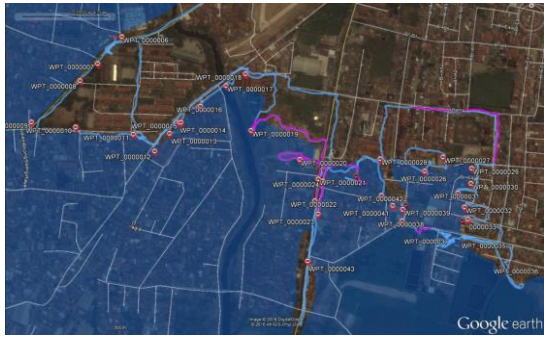

**Fig.3.** Flood plotting results on a segment in the field while the team conducted a flood inundation survey on 25 December 2014.

Incidentally, flood inundation mapping on 13 March 2016 coincides with the capture time of satellite image, documented on Google Earth servers. Accordingly, we can compare mapping flood conditions to mapping results and flood image imagery from Google Earth. **Fig.4** shows the comparison between the two maps. In a qualitative way, we can conclude that the rapid mapping method could represent the flooded submerged area in the residential area, but it is less likely to represent the flooded area in the rice field location. This shortcoming is caused by the transportation used by the surveyor team is motorcycle, so it can not reach the rice field area.

Table 1. The speed, distance, and time range of the flood survey using the **Orux Map**.

| Teams | Year | Time range (h:m) | Distance (km) | Average speed (km/h) |
|---|---|---|---|---|
| Team 1 | 2014 | 05:04 | 19.4 | 3.8 |
| Team 2 | 2016 | 01:48 | 14.3 | 7.9 |
| Team 3 | 2016 | 03:23 | 2.5 | 4.6 |
| Team 4 | 2016 | 01:39 | 14.3 | 7.9 |
| Team 5 | 2016 | 01:46 | 11.9 | 6.8 |

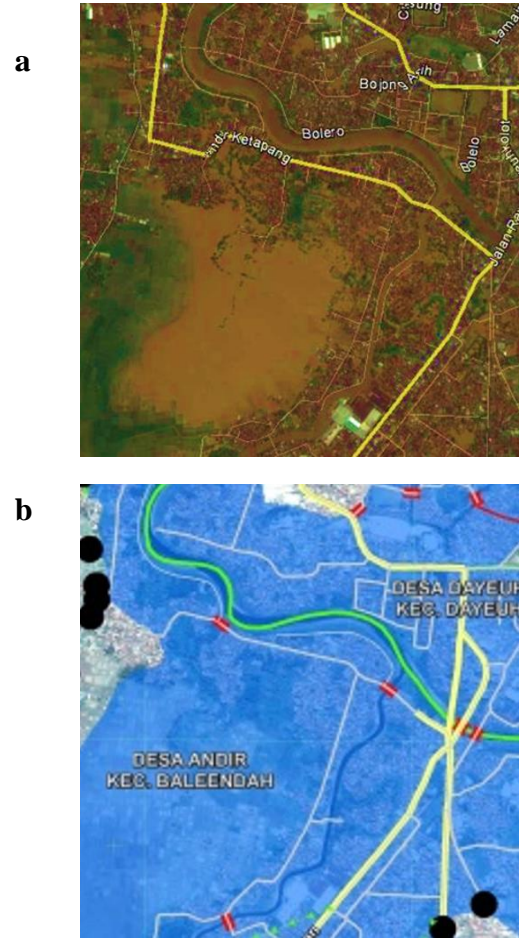

**Fig.4.** Comparison of flood inundated areas in Baleendah (a) the result of Google Earth satellite imagery and (b) mapping results by rapid flood survey method on 13 March 2016.

## 4 Conclusion

The free Android-based mapping application, **Orux Map** can be used to perform flood mapping quickly, cheaply and accurately and can be utilized to support emergency response. Survey speeds averaged about 6.2 km/h, with a survey team consist of two surveyor personnel; one motorcycle; one smartphone with Android operating system; and internet quota. The mapping results show compatibility with satellite imagery and aerial photographs. The flooded areas in Baleendah and surrounding areas on 25 December 2014 is 794 Ha and on 13 March 2016 is 794 Ha. In addition, this mapping method is easy to learn and use by local communities.

**Acknowledgements**

First and foremost, we are extremely grateful to the local surveyor participants from Garda Caah and Jaga Balai communities, who took the time from their busy schedules to participate in the study. Without their participation and feedback, this study would not have been possible. We thank Abdul Haris Wirabrata for comments that greatly improved the manuscript. Funding for this research, carried out by Atmospheric Sciences Research Group ITB, was provided by the Institute for Research and Community Services (LPPM) ITB as part of the Bandung Basin Flood Mapping Project.